\documentclass[aps,prl,twocolumn,showpacs,superscriptaddress]{revtex4}
\usepackage[dvips]{graphicx}
\usepackage{amssymb}
\usepackage{amsmath}
\usepackage{color}
\usepackage{xspace}
\begin{document}

\title{Counting Statistics in an InAs Nanowire Quantum Dot with a Vertically Coupled Charge Detector}

\author{T. Choi}
\affiliation{Solid State Physics Laboratory, ETH Zurich, 8093
Zurich, Switzerland}
\author{T. Ihn}
\affiliation{Solid State Physics Laboratory, ETH Zurich, 8093
Zurich, Switzerland}
\author{S. Sch$\mathrm{\ddot o}$n}
\affiliation{FIRST lab, ETH Zurich, 8093 Zurich, Switzerland}
\author{K. Ensslin}
\affiliation{Solid State Physics Laboratory, ETH Zurich, 8093
Zurich, Switzerland}
\date{\today}

\date{\today}

\begin{abstract}
A gate-defined quantum dot in an InAs nanowire is fabricated on top of a quantum point contact realized in a two-dimensional electron gas. The strong coupling between these two quantum devices is used to perform time-averaged as well as time-resolved charge detection experiments for electron flow through the quantum dot. We demonstrate that the Fano factor describing shot noise or time-correlations in single-electron transport depends in the theoretically expected way on the asymmetry of the tunneling barriers even in a regime where the thermal energy $k_\mathrm{B}T$ is comparable to the single-particle level spacing in the dot.
\end{abstract}

\maketitle

Single spins in semiconductor quantum dots (QDs) are considered as possible candidates for qubits for solid-state quantum information processing \cite{loss98}. Many essential experiments have been done in split-gate defined QDs formed in GaAs/AlGaAs two-dimensional electron gases (2DEGs) where coherent rotations of single spins and the coherent exchange of two spins have succesfully been demonstrated \cite{koppens06,petta05}. While the realization of a spin qubit in such split-gate defined QDs in GaAs/AlGaAs 2DEGs is well established by now, much work focusses on the implementation of spin qubits in other material systems, where different material properties could be advantageous for both longer coherence times or fast manipulation of the spin. For example, systems with negligible nuclear spin and weak spin-orbit interaction such as carbon nanotubes \cite{churchill09}, graphene \cite{trauzettel07}, or Si-based systems \cite{morello10,simmons11} are thought to be promising due to the expected long spin coherence time. On the other hand, systems with strong spin-orbit interaction would promise an efficient manipulation of the spin using only electric fields \cite{golovach06,flindt06}. Spin-orbit mediated coherent rotation of a single spin has been demonstrated in split-gate defined GaAs QDs \cite{nowack07} and recently, in an InAs nanowire establishing a so-called spin-orbit qubit \cite{nadjperge10}.

For this purpose, InAs is an interesting material due to the small effective mass of the electrons ($m^{\star}=0.023m_{0}$) leading to large confinement energies, the large effective $g^{\star}$-factor and the strong spin-orbit interaction. In \cite{nadjperge10}, the QDs were formed by thin metallic gates lying below the nanowire and the spin states were measured by direct transport through the nanowire using spin-to-charge conversion. A less invasive way to measure the charge on a QD is to use charge detection by a nearby quantum point contact (QPC) \cite{field93}. However, due to the given geometry, it is not straightforward to implement a sensitive charge detector for a nanowire QD \cite{Hu07,shorubalko08,choi09}. 

Here, we present a method to fabricate top-gate defined QDs in an InAs nanowire with a charge detector lying exactly below the nanowire. The top gate technique ensures a high tunability like the samples in \cite{nadjperge10, pfund06} which, however, did not include a charge detector. The design of the charge detector results in strong coupling between the QD and the detector like in \cite{shorubalko08, choi09}, but improving on the limited tunability in previous devices.

\begin{figure}[h]
\centerline{\includegraphics[angle=0,width=8.5cm,keepaspectratio,clip]{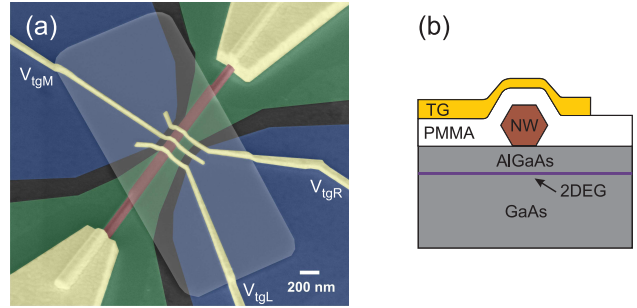}}
\caption{\textbf{a)} Tilted SEM image of a sample with the same design as the one measured (false colors). The InAs nanowire (in
red) is deposited on an AlGaAs/GaAs heterostructure with a
two-dimensional electron gas (2DEG) 37 nm below the surface. Etched trenches in the 2DEG define a QPC (in green) below the nanowire. Side gates (in blue) are used to tune the QPC. Three top gates (in yellow) on top of the crosslinked PMMA (bright pad) are used to define the QDs. \textbf{b)} Schematical cross section of the sample. The PMMA serves as a gate insulator between the nanowire and the top gates.}\label{figure1}
\end{figure}

The InAs nanowires used in this work are grown by metal organic vapour-phase epitaxy on a $\langle 111 \rangle$B oriented GaAs substrate using colloidal Au particles as catalysts \cite{pfund06chimia}. The crystal structure of the nanowires is wurtzite and the nanowires are typically $100~\mathrm{nm}$ in diameter and up to $10~\mu\mathrm{m}$ long. The nanowires have a hexagonal cross section and grow perpendicular to the substrate. After growth, the nanowires are deposited on a molecular beam epitaxy grown AlGaAs/GaAs heterostructure containing a two-dimensional electron gas (2DEG) $37~$nm below the surface, where the nanowires come to lie parallel to the surface of the heterostructure. The 2DEG has an electron density of $N_{\mathrm{s}}=4 \times 10^{11}~\mathrm{cm}^{-2}$ and a mobility of $3 \times 10^{5}~\mathrm{cm}^{2}/\mathrm{Vs}$ at $T=2~\mathrm{K}$. Ti/Au ohmic contacts for the nanowire are fabricated using electron beam lithography (EBL) and metal evaporation. In order to remove the native oxide of the nanowire prior to the ohmic contact deposition, a single-step etching/passivation procedure with a diluted ammonium polysulfate ((NH$_{4}$)$_{2}$S$_{x}$) solution is used \cite{suyatin07}. The QPC charge detector is defined by wet chemical etching in such a way that the detector is positioned exactly below the nanowire. A pad of crosslinked PMMA \cite{zailer96} on top of the nanowire is used as a gate insulator. As a final step, three top gates are fabricated by EBL and the evaporation of Ti/Au. The top gates have a width of $40~$nm and a spacing of $128~$nm. A false color scanning electron micrograph (SEM) image and a schematical cross section of the sample are shown in Fig.~\ref{figure1}. The QPC (in green) acts as a charge detector for the nanowire QDs and at the same time as a global back gate to tune the electron density in the nanowire. The side gates in the 2DEG (in blue) are used to tune the conductance of the QPC to a slope where it is sensitive to changes of the electron number of the QDs. All measurements presented are performed at a temperature of $T=1.5~$K.

\begin{figure}[h]
\centerline{\includegraphics[angle=0,width=8.5cm,keepaspectratio,clip]{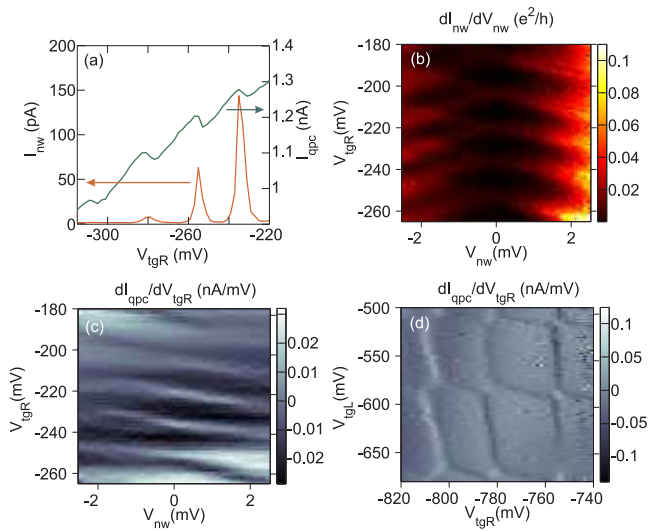}}
\caption{ \textbf{a)} Current $I_{\textrm{nw}}$ through the QD (in red) and simultaneous measurement of the QPC (in green). A leakage current of $11~$pA from the 2DEG to the ohmic contacts of the nanowire has been substracted from $I_{\textrm{nw}}$. Steps in $I_{\textrm{qpc}}$ can be seen at the positions of the Coulomb peaks in $I_{\textrm{nw}}$. The bias across the QPC is $V_\textrm{qpc}=100~\mu$V. \textbf{b)} Coulomb diamonds in the differential conductance $\textrm{d}I_{\textrm{nw}}/\textrm{d}V_{\textrm{nw}}$. \textbf{c)} Numerical transconductance $\mathrm{d}I_\mathrm{qpc}/\mathrm{d}V_\mathrm{tgR}$ of the QPC measured simulteneously. \textbf{d)} Charge stability diagram in the DQD regime measured by charge detection. $V_\textrm{nw}=0~$mV.}{\label{figure2}}
\end{figure}

By applying negative voltages to the top gates, QDs can be formed in the nanowire. In Fig.~\ref{figure2}(a) clear Coulomb blockade peaks can be seen in the current through the nanowire (red). At the same time, the QPC is tuned to a sensitive operating point using the side gates. Steps in the conductance of the QPC appear whenever the electron number of the QD changes by one (green). The QPC is able to detect transitions even when the current through the nanowire gets too small to be measured directly, as can be seen for $V_\mathrm{tgR}=-307~$mV. A measurement of the differential conductance $dI_\mathrm{nw}/\textrm{d}V_{\textrm{nw}}$ against the bias voltage $V_\mathrm{nw}$ across the QD and $V_\mathrm{tgR}$ is shown in Fig.~\ref{figure2}(b). The top gates have been set to values where a large single QD is formed between the left and the right outer top gates. The middle top gate is set to $V_\mathrm{tgM}=-50~$mV. Coulomb diamonds are seen from which a charging energy of $E_{\mathrm{c}}\approx 1~$meV is deduced. We assume the QD to be a prolate ellipsoid with capacitance $C_{\Sigma}=4 \pi \epsilon_{0} \epsilon \sqrt{a^{2}-b^{2}}/\mathrm{ln}\left(a/b+\sqrt{(a/b)^{2}-1}\right)$, where $\epsilon=15$ for InAs, $a$ is the semi-major axis, and $b$ is the semi-minor axis. With $E_{\mathrm{c}}=e^{2}/C_{\Sigma}$ and setting $2b=93~$nm for the diameter of the nanowire as measured by SEM, a value of $2a \approx 220~$nm is obtained. This is reasonable for a large single QD lying between the outer two top gates. These numbers give a rough estimate of $\Delta E \sim 70~\mu$eV for the single-particle level spacing, which is comparable to $k_\mathrm{B}T$.

Figure~\ref{figure2}(c) shows a measurement of the transconductance $\mathrm{d}I_\mathrm{qpc}/\mathrm{d}V_\mathrm{tgR}$ of the QPC measured simultaneously with the differential dot conductance. Changes in electron number can be detected, but in contrast to the measurement of $I_\mathrm{nw}$ in Fig.~\ref{figure2}(b), only lines with negative slopes are visible. This indicates asymmetric coupling of the QD states to the leads \cite{rogge05}. In the measured setup, the lines with negative slope correspond to the situation where the electrochemical potential of the QD is aligned with the Fermi level in the drain ($\mu_\mathrm{N}=\mu_\mathrm{D}$). The QPC monitors the average charge on the QD which, in case of asymmetric barriers, is determined by the barrier with the higher tunneling rate. Thus, in the case of Fig.~\ref{figure2}(c), the tunneling rate to the drain is much higher than the tunneling rate to the source ($\Gamma_\mathrm{D} \gg \Gamma_\mathrm{S}$). 

By tuning the middle top gate to more negative values, a double quantum dot (DQD) is formed in the nanowire. Figure~\ref{figure2}(d) shows a measurement of the transconductance $\mathrm{d}I_\mathrm{qpc}/\mathrm{d}V_\mathrm{tgR}$ of the QPC versus the two outer top gates $V_\mathrm{tgL}$ and $V_\mathrm{tgR}$ at a middle top gate value $V_\mathrm{tgM}=-650~$mV. The characteristic honeycomb diagram expected for DQDs can be recognized \cite{vanderwiel03}.

\begin{figure}[h]
\centerline{\includegraphics[angle=0,width=8.5cm,keepaspectratio,clip]{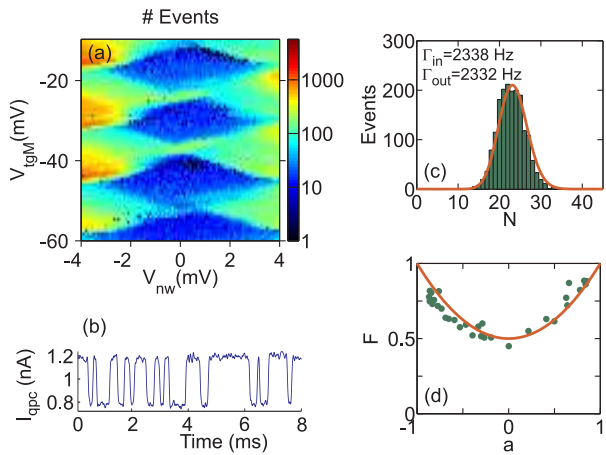}}
\caption{ \textbf{a)} Coulomb diamonds measured by time-resolved charge detection. A $1~$s time trace is taken at each point of the plot. $V_\mathrm{tgR}=-955.3~$mV, $V_\mathrm{tgL}=-942.6~$mV, $V_\mathrm{qpc}=250~\mu$V \textbf{b)} Typical time trace of the QPC current, lowpass filtered at $20~$kHz. $V_\mathrm{qpc}=100~\mu$V \textbf{c)} Distribution of the electrons tunneling through the QD for fixed $\Gamma_\mathrm{S/D}$. The solid line is the theoretical distribution \cite{bagrets03}. \textbf{d)} Fano factor measured at different asymmetries. The solid line is the theoretical prediction $F=(1+a^2)/2$.}{\label{figure3}}
\end{figure}

Another advantage of using a QPC as a charge detector for a QD is the possibility to perform time-resolved detection of single electrons passing through the QD \cite{schleser04, vandersypen04}. In particular, it provides the possibility to measure the full counting statistics (FCS) and offers thus a powerful tool to investigate the noise properties of the QD \cite{levitov96, bagrets03, gustavsson06}. For this purpose, the device is again tuned to the single dot regime with $V_\mathrm{tgL}$ and $V_\mathrm{tgR}$ set to values where the tunneling rates $\Gamma_\mathrm{S/D}$ get smaller than the experimental bandwidth of $\sim 33~$kHz. The middle gate is used as a plunger gate to change the electron population of the QD. In this regime, a time trace of the QPC conductance shows switching between two distinct levels for $N$ or $N+1$ electrons being on the QD. A typical time trace is shown in Fig.~\ref{figure3}(b). The relative change of the QPC conductance is $\Delta G_{\mathrm{QPC}}/G_{\mathrm{QPC}} \approx 30$\% and thus much larger than values of $1$\%$-2$\% for split-gate defined QDs. In other regimes, $\Delta G_{\mathrm{QPC}}/G_{\mathrm{QPC}}$ can even be up to $60$\%. Figure~\ref{figure3}(a) shows a measurement of Coulomb diamonds where a $1~$s time trace is taken at each point of the plot and the number of events, where a single electron tunnels from the QD to any lead, is counted. At a fixed bias $V_\mathrm{nw}\gg k_\mathrm{B} T$ and with the electrochemical potential $\mu_{\mathrm{N}+1}$ of the QD in the bias window, the electron transfer through the QD happens always from source to drain without tunneling back. In this case, the current through the QD is equal to the number of electrons transferred through the QD and the probability distribution function $p_{t_0}(N)$ of $N$ electrons passing through the QD in a time interval $t_0$ can be measured. Such a distribution at fixed $\Gamma_\mathrm{S}$ and $\Gamma_\mathrm{D}$ is shown in Fig.~\ref{figure3}(c). The length of the time intervals is $t_0=20~$ms and the bias across the QD is $V_\mathrm{nw}=2~$mV. The red solid line is the theoretical solution for the probability distribution function given by
\begin{equation}
p_{t_0}(N)=\frac{1}{2\pi}\int_{-\pi}^\pi e^{-i N \chi -S(\chi)}d\chi
\label{eq:pN}
\end{equation}
with $S(\chi)=\frac{t_0}{2}[\Gamma_\mathrm{S}+\Gamma_\mathrm{D}-\sqrt{(\Gamma_\mathrm{S}-\Gamma_\mathrm{D})^2+4\Gamma_\mathrm{S}\Gamma_\mathrm{D}e^{i\chi}}]$ the generating function and $e^{i\chi}$ the counting field \cite{bagrets03}. We attribute the small deviation of the data with respect to the model to the occasional presence of three-level traces, which could arise from two excess electrons tunneling through the QD. This is reasonable considering the small charging energy of the QD and the rather high temperature of $T=1.5~$K.

Since electron transport through a QD is governed by Coulomb blockade, the noise is expected to be sub-Poissonian. Thus the Fano factor $F=S_I/2eI$, with $S_I$ the shot noise and $I$ the average current, is smaller than one. For a QD, one gets for the Fano factor $F=(1+a^2)/2$, with $a=(\Gamma_\mathrm{S}-\Gamma_\mathrm{D})/(\Gamma_\mathrm{S}+\Gamma_\mathrm{D})$ the asymmetry of the tunnel barriers \cite{bagrets03}. The Fano factor $F=\langle (N-\langle N \rangle)^2 \rangle / \langle N \rangle$ is extracted from the width and the mean of the experimental distributions like that in Fig.~\ref{figure3}(c). Figure~\ref{figure3}(d) shows the Fano factor measured for different values of $a$, where $\Gamma_\mathrm{S/D}$ was tuned using $V_\mathrm{tgL/R}$. The measured points are in good agreement with the theoretical prediction over the whole range of $a$.

In conclusion, we have presented a design for highly tunable InAs nanowire QDs with very sensitive charge detection. The high tunability and sensitivity enabled time-resolved charge detection and the measurement of the FCS. The data agrees with the FCS theory for single-level transport even though $k_\mathrm{B}T$ is comparable to the single-particle level spacing in the dot. By reducing the size of the QDs, we expect to be able to reach the few-electron regime and to carry out single-spin detection and manipulation with the help of a charge detector.

We thank Y.~Komijani, B.~K$\mathrm{\ddot u}$ng, T.~M$\mathrm{\ddot u}$ller, and A.~Alt for valuable help and discussions, and I.~Shorubalko and E.~Gini for supervision in nanowire growth.

\end{document}